\documentclass[prd,nofootinbib,floats,aps,onecolumn]{revtex4}
\usepackage{graphicx}
\usepackage{amssymb}
\usepackage{amsmath}
\usepackage{color}
\usepackage{bm}
\usepackage[hypertex]{hyperref}
\usepackage{amsbsy}
\usepackage{epsfig}

\newcommand{\half}{\frac{1}{2}}

\newcommand{\be}{\begin{equation}}
\newcommand{\ee}{\end{equation}}
\newcommand{\bea}{\begin{eqnarray}}
\newcommand{\eea}{\end{eqnarray}}
\newcommand{\ben}{\begin{enumerate}}
\newcommand{\een}{\end{enumerate}}
\newcommand{\bit}{\begin{itemize}}
\newcommand{\eit}{\end{itemize}}
\newcommand{\eq}[1]{eq.~(\ref{#1})}

\newcommand{\la}[1]{\label{#1}}
\newcommand{\Eq}[1]{Eq.~(\ref{#1})}

\newcommand\brak[1]{\ensuremath{\bigl| #1\bigr\rangle}}
\newcommand\krab[1]{\ensuremath{\bigl\langle #1\bigr|}}
\newcommand\brakket[2]{\ensuremath{\,\bigl\langle #1 \big| #2
\bigr\rangle\,}}

\def\c2{$\chi_2$}

\def\sch{Schr\"odinger }

\newcommand{\pvalt}{\raise0.15ex\hbox{-}\mkern-11.5mu\int}
\newcommand{\llll}{ \ell (\ell + 1)}

\definecolor{BrickRed}{cmyk}{0,0.89,0.94,0.28}
\definecolor{MidnightBlue}{cmyk}{0.98,0.13,0,0.43}
\definecolor{DarkGreen}{rgb}{0,0.7,0.1}
\begin{document}

\preprint{MIT-CTP 3802}
\title{Ordinary and Extraordinary Hadrons\footnote{Invited talk presented at the 2006 Yukawa International Seminar {\sl New Frontiers in QCD}, Kyoto University, Kyoto, Japan, November, 2006.}}

\author{R.~L.~Jaffe}
\affiliation{Center for Theoretical Physics,\\
Laboratory for Nuclear Science and Department of Physics\\
Massachusetts Institute of Technology, \\Cambridge, Massachusetts
02139  }

\begin{abstract} 
Resonances and enhancements in meson-meson scattering can be divided into two classes distinguished by their behavior as the number of colors ($N_{c}$) in QCD becomes large: The first are \emph{ordinary} mesons that become  stable as $N_{c}\to\infty$.  This class includes   textbook $\bar q q$ mesons as well as glueballs and hybrids.  The second class, \emph{extraordinary} mesons, are enhancements that disappear as $N_{c}\to \infty$; they subside into the hadronic continuum.  This class includes indistinct and controversial objects that have been classified as $\bar q \bar q qq$ mesons or meson-meson molecules.  Pel\'aez's study of the $N_{c}$ dependence of unitarized chiral dynamics illustrates both classes: the $p$-wave $\pi\pi$ and $K\pi$ resonances, the $\rho(770)$ and $K^{*}(892)$, behave as ordinary mesons; the $s$-wave $\pi\pi$ and $K\pi$ enhancements, the $\sigma(600)$ and $\kappa(800)$, behave like extraordinary mesons.  Ordinary mesons resemble Feshbach resonances while    extraordinary mesons look more like effects due to potentials in meson-meson scattering channels.  I build and explore toy models along these lines.  Finally I discuss some related dynamical issues affecting the interpretation of extraordinary mesons.

\end{abstract}

%
%
\maketitle

\begin{center}
{\sl In memory of R. H. Dalitz}
\end{center}

\section{Introduction}
Dick Dalitz, together with his students and collaborators, made many fundamental contributions to our understanding of   hadron-hadron interactions. In addition to the famous  Dalitz plot\cite{dalitzplot}, Dalitz, together with Castellejo and Dyson, showed how elementary particles would appear in unitary descriptions of low energy scattering --- the so-called \emph{CDD poles}\cite{Castillejo:1955ed} --- and complicate attempts to implement the bootstrap program.  Later Dalitz and Tuan pioneered the use of $K$-matrix methods to extract information on hyperon resonances from low energy two-body scattering\cite{Dalitz:1960du}. 

The analysis of low energy hadron-hadron scattering remains problematic and controversial to this day.  Of course \emph{weakly coupled}, narrow resonances like the $J/\psi(3097)$ or the $\phi(1020)$ pose no problem.  They are can be identified with $S$-matrix poles just below the real axis in the complex energy plane.  Their lifetimes are long compared to typical strong interaction timescales, so they can be regarded as approximate stationary states of the QCD Hamiltonian, and their properties can be computed without paying much attention to the open channels into which they eventually decay.

However most (light quark) hadronic resonances are several hundred MeV wide and overlap one another.    Some, like the $f_{0}(600)$ (the ``$\sigma$'' of the $\sigma$-model),   are so broad that their resonant character has only recently been confirmed\cite{Caprini:2005an}.  Others, like the enhancement observed in low energy $s$-wave $K\pi$ scattering (known as either the $K^{*}_{0}(800)$ or the $\kappa$), have not yet been convincingly identified with a pole in the $S$-matrix\cite{kappa}.   It is not obvious how to interpret these enhancements in terms of the underlying quark and gluon degrees of freedom of QCD.  This talk lays out some qualitative ideas on the nature of enhancements, both narrow and broad, in QCD, where permanently confined, kinematically closed, and open channels coexist for most two body scattering channels.  

Discoveries on several fronts have recently re-envigorated the study of hadron resonances\cite{Rosner:2006jz}.  New, unusual charmonium states, like the $X(3872)$ and $Y(4260)$ have been interpreted as charm-molecules, four-quark states, or hybrids (quark-gluon composites).  The discovery of charm-strange mesons with unexpected masses have stimulated similar speculations.  On another front, there continues to be considerable interest in the long-standing problem of the light scalar mesons, much of it suggesting that the lightest ($f_{0}(600)$, $K^{*}_{0}(800)$, $f_{0}(980)$, and $a_{0}(980)$) scalar mesons are not dominantly  ordinary $ \bar q q$ states, though there is little agreement on what they actually \emph{are} \cite{Close:2002zu,Amsler:2004ps,Achasov:2006sr,Yao:2006px,vanBeveren:2006ua,Anisovich:2005kt}.  The tools that are used to analyze these systems --- relativistic and non-relativistic quark models, $K$-matrix and other unitarization schemes, and QCD-sum rules, for example --- work reasonably well for ordinary $\bar q	q$ mesons, but seem to be too blunt to give definitive answers for these unusual states.   Debates rage, for example, between proponents of molecules, $(\bar qq)(\bar qq)$, and four quark states, $\bar q \bar q qq$, when in fact the distinction between these two classes of states has never been made clear.

The principal aim of this talk is to describe a division of mesons 
into two classes, ``\emph{ordinary}'' and ``\emph{extraordinary}''.   The  distinction between the classes is a theoretical one:  ordinary mesons decouple as the number of colors ($N_{c}$) in QCD becomes infinite.  Their widths go to zero; they become stable.  In contrast, extraordinary mesons disappear as $N_{c}\to\infty$; they subside into the hadronic continuum.  Similar ideas may apply to baryons, which have different large $N_{c}$ systematics, but I do not discuss them here.  The distinction between ordinary and extraordinary is motivated by general large $N_{c}$ counting arguments.  It is supported in one particularly important case by recent work by Jos\'e Pelaez on the $N_{c}$ dependence of the chiral Lagrangian and its predictions for low energy $\pi\pi$ scattering\cite{Pelaez:2003dy}.  

\emph{Ordinary mesons} include the traditional $\bar q q$ mesons, hybrids, and glueballs of quark models.  They decouple from scattering channels as $N_{c}\to\infty$, and become \emph{bound states in the continuum}.  As described in \S2, it may be useful to think of them as   \emph{Feshbach resonances}.  

\emph{Extraordinary mesons} include multiquark states, exotics, and meson-meson molecules.  As $N_{c}\to\infty$ these ``states'' just go away.  From the point of view of low-energy two-body scattering, they resemble effects generated by an $s$-channel potential, described, for example, as zeros in the denominator in the old $N/D$ method\cite{chewmandelstam,frautschi}.  A leading candidate for this kind of hadron is the $\sigma$ or $f_{0}(600)$, which is the chiral partner of the pion and the QCD analogue of the Higgs boson in technicolor theories.  $\sigma$ exchange contributes to the intermediate range attraction that  binds  nuclei.    The fact that this important meson disappears as $N_{c}\to\infty$ is a reminder that there are some crucial aspects of QCD that are lost in this limit.  Diquark correlations, well known to be important in QCD spectroscopy, also disappear as $N_{c}\to\infty$\cite{diquarks}.
  
This talk is organized as follows:  Section 2 is devoted to ordinary mesons.  The rules for large $N_{c}$ counting are familiar from textbooks.  They  predict  that ordinary mesons, $\bar qq$ for example, should decouple as $N_{c}\to\infty$.  Sadly, only theorists can vary $N_{c}$, and no theoretical framework for studying hadron-hadron scattering in QCD is robust enough to test this prediction in general.  Low energy pseudoscalar meson-meson scattering is a fortunate exception.  The $N_{c}$ dependence of the chiral Lagrangian has been established\cite{chiral}, and with the help of constraints from $s$-channel unitarity, it can be used to analyze the $N_{c}$ dependence of the low energy enhancements in $\pi\pi$ and $K\pi$ scattering.  Pel\'aez's analysis of the $\pi\pi$ and $K\pi$ $p$-waves, shows that the resonances in these channels, the $\rho$ and $K^{*}$ mesons, behave just as large $N_{c}$ predicts:  They decouple as $N_{c}\to\infty$.  

When $N_{c}$ is very large and ordinary mesons are very narrow, the space of meson states in QCD factors neatly into two sectors:  Permanently confined channels with a discrete states, and the meson-meson continuum, where interactions are, in general, ${\cal O}(1/N_{c})$.  The discrete states develop widths of ${\cal O}(1/N_{c})$ due to their weak coupling to the continuum.  An analogous situation is well-known in atomic and nuclear physics, where the narrow resonances, interpreted as bound states in the continuum, are known as \emph{Feshbach} resonances\cite{feshfano}. In Section 2 I review the Feshbach resonance formalism and build a toy model to describe the $N_{c}$ dependence of ordinary hadrons in this formalism.

Section 3 introduces extraordinary hadrons.  The generic large $N_{c}$ analysis is an extension of Ref.~\cite{Jaffe:1981up}.  $N_{c}$ counting suggests that the leading (${\cal O}(1/N_{c})$) contribution to meson-meson scattering \emph{off-resonance} comes from quark exchange diagrams that do not couple to confined channels.  These diagrams may give rise to important meson-meson interactions, even though they should disappear as $N_{c}\to\infty$.  Once again unitarized chiral perturbation theory provides a case in point.  Pel\'aez's analysis of the low energy $\pi\pi$ and $K\pi$ $s$-wave suggests that the $f_{0}(600)$ and $\kappa$ mesons are examples of extraordinary hadrons that disappear at large $N_{c}$\cite{Pelaez:2003dy}.  It is important to emphasize that these \emph{extraordinary hadrons} are not necessarily manifested as resonances.  There may be no pole in the $S$-matrix near enough to the physical region to identify unambiguously with a particular enhancement in the data.  After all, as $N_{c}\to\infty$, the effects must \emph{go away}, meaning that the singularities in the $S$-matrix must recede into the distant complex plane. 

The enhancements I identify as extraordinary mesons  cannot be generated by the exchange of ordinary  mesons, nor are they in any sense dual to the narrow, ordinary, $s$-channel resonances.  Instead they look ``\emph{potential-like}'' --- in non-relativistic scattering they would be described by a potential a  \sch equation with only open (or perhaps kinematically closed) channels.  They do not communicate with the confined channels where ordinary meson resonances live.  One relativistic generalization of non-relativistic potential scattering is provided by the $N/D$ method of Chew and Mandelstam\cite{chewmandelstam}.  After introducing the $N/D$ method I build a toy model of extraordinary mesons based on a separable potential that illustrates the fact that they disappear as the interaction strength vanishes.  Although this is obvious without the model, it is interesting to contrast the dependence of ordinary and extraordinary hadrons on $N_{c}$.  The results are shown in Figs.~(\ref{compares}) and (\ref{comparep}), which emphasize the distinction between the two classes of objects.

 Section 4 contains a collection of comments on the analysis of low energy hadron-hadron scattering and the interpretation of extraordinary hadrons.  First, I point out that the meson-meson $s$-wave may be uniquely sensitive to extraordinary hadrons.  Second, I review the arguments that the eigenstates of model QCD Hamiltonians computed with confining interactions or boundary conditions do not, in general, correspond to ordinary hadrons.  For example, a model QCD Hamiltonian will have a tower of eigenstates with $uu\bar d \bar d$ quantum numbers if the coupling to the open $\pi^{+}\pi^{+}$ decay channel is blocked.  However, these are \emph{not}  resonances.  Quite the contrary, they may be evidence for  \emph{repulsive} interactions in that channel.   It is more correct to view such calculations as   sources of information about  the interactions that may generate extraordinary hadrons.  Third I point out that the distinction between a meson-meson molecule and a $\bar q\bar q q q$ state is unclear, and depends on what question is being asked.  Finally I review the physical significance of $K$-matrix poles and caution against identifying them with QCD ``states'' when their residues are large.

\section{Ordinary mesons}
\label{largn}

Ordinary mesons are the quark-antiquark, glueball, and hybrid states discussed in all textbooks on QCD.  They have in common the property that they cannot decay until other quanta --- $\bar qq$ pairs or gluons --- are created by interactions.  This makes them relatively narrow and, if narrow, relatively straightforward to describe poles  in scattering amplitudes nearby the physical region.  Here, as a tune-up, I review the large-$N_{c}$ analysis of ordinary mesons.  Next, I show that chiral dynamics supports the interpretation of the light vector mesons like the $\rho$ and $K^{*}$ as ordinary mesons.   Finally I introduce Feshbach resonances, which are not very well known among particle physicists, and point out their resemblance to ordinary mesons.

\subsection{Large $N_{c}$ expectations}

The behavior of meson resonances in the large $N_{c}$ hardly needs discussion.  The limit was first  
worked out by 't Hooft\cite{'tHooft:1973jz}, and had been nicely reviewed, for example, by Coleman\cite{Coleman:1980nk}.  For us, it suffices to remember that $\bar q q$-meson wave functions, quark-quark-gluon and three gluon vertices carry   factors of $1/\sqrt{N_{c}}$, four gluon vertices scale like $1/N_{c}$, and quark loops contribute a factor of $N_{c}$.  This is enough to show that the amplitudes for $M\to M_{1}M_{2}$ (all $\bar q q$ mesons) are proportional to $1/\sqrt{N_{c}}$, and therefore that meson widths vanish like $1/N_{c}$ as $N_{c}\to\infty$.  Mixing of higher Fock components, $\bar q\bar q q q, \bar q\bar q\bar qqqq,...$, into $\bar q q$ meson states is suppressed by powers of $N_{c}$.  So $\bar q q$ mesons have definite quark and antiquark numbers as $N_{c}\to\infty$.  Glueball and hybrid wavefunctions carry factors of $1/N_{c}$. Using the $N_{c}$ scaling of loops and vertices, it is easy to show that glueball and hybrid widths also
vanish as $N_{c}\to\infty$.  Although glueballs and hybrids are far from ordinary in other ways, they decouple from scattering channels as $N_{c}\to\infty$ and belong in the ``ordinary'' category for the present discussion.

\subsection{Evidence from chiral dynamics}
\label{chiralpwave}

As $N_{c}\to\infty$, meson-meson scattering should simplify dramatically.  There should be no interaction at all except in the immediate vicinity of ordinary meson resonances, whose  widths are ${\cal O}(1 /N_{c})$.
Can this behavior be verified by direct computation of the $N_{c}$ dependence of meson-meson scattering in QCD?  Unfortunately, there is almost no first-principle   information on hadron scattering in QCD.  An important exception is $\pi\pi$ scattering at low energies, where chiral dynamics provides an effective Lagrangian as a power series in $ p^{2}/\Lambda_{\chi}^{2}$ (where $\Lambda_{\chi}$ is the scale of spontaneous chiral symmetry violation).  It turns out that there is much to be learned from this case.  In the limit of exact $SU(2)_{L}\times SU(2)_{R}$ the only parameter at ${\cal O}(p^{2})$ is $f_{\pi}$, the pion decay constant.  At the next order (${\cal O}(p^{4}))$ eight parameters, $L_{1} ... L_{8}$, enter\cite{Gasser:1983yg}.  Over the years the values of these parameters have been estimated and their leading $N_{c}$ dependence has been established\cite{chiral}.  Of course a finite number of terms in an expansion in $p^{2}$ cannot generate a pole in a scattering amplitude.  Unitarity adds important constraints that guide the extrapolation of low energy $\pi\pi$ scattering to larger $p^{2}$ where enhancements and poles do appear.  The ``inverse amplitude'' method (IAM), originally developed by Truong\cite{truong} and later adapted to meson-meson scattering in chiral pertrubation theory\cite{iam}, provides an explicit and well-developed example.  In the case of a single channel it is particularly simple:  unitarity determines the imaginary part of the inverse of the partial wave scattering amplitude uniquely.  The $\pi\pi$ scattering amplitude with isospin $I$ and angular momentum $J$ obeys
\be
\la{inverse}
t_{IJ}^{-1}(p)=   g_{IJ}(p^{2}) -\frac{ip}{\sqrt{p^{2}+m_{\pi}^{2}}}.
\ee
Here $p$ is the pion momentum in the $\pi\pi$ center of mass, so $s=4(p^{2}+m_{\pi}^{2})$.  $t_{IJ}(p)$ is related to the \emph{Argand amplitude}, $f_{IJ}(p)$, by
$$
t_{IJ}(p)=\frac{\sqrt{p^{2}+m_{\pi}^{2}}}{p}f_{IJ}(p),\quad \mbox{where}\quad f_{IJ}(p)
=\sin\delta_{IJ}(p)e^{i\delta_{IJ}(p)}.
$$
Chiral perturbation theory is used to expand $g_{IJ}$ about $p^{2}=0$. \Eq{inverse}  can then be extrapolated to larger values of $p$.  The extrapolation can be used to extract the $\pi\pi$-phase shift, or to identify a resonance if a zero in $t_{IJ}^{-1}$ appears at not too large a value of $p$, not too far from the real $p$-axis.  How reliable are such extrapolations?  Clearly they are only as reliable as the order to which $g_{IJ}(p^{2})$ is computed since it is not possible to unambiguously identify a pole in an analytic function from any finite number of terms in a Taylor expansion.\footnote{Recently Pel\'aez and Rios have carried the calculation of $g_{IJ}$ to two loop order and confirmed earlier results \cite{Pelaez:2006nj}.  Also, their earlier results have been qualitatively confirmed by  
Uehara\cite{Uehara:2003ax}.}   In addition one might expect that several different unitary scattering  amplitudes have the same inverse real part, {\it i.e.\/} that the unitarization method is ambiguous.  It can be shown, however, that when only a single channel is open, the inverse amplitude method is unique\cite{Pelaez:2006qv}.  This is good enough for the study of low energy $\pi\pi$ and $K\pi$ scattering including the regions of the $\sigma$, $\kappa$, $\rho$, and $K^{*}$.    In any case, we are interested in the \emph{qualitative} behavior as a function of $N_{c}$, not in quantitative details.  

The application of these methods to $\pi\pi$ and $K\pi$ scattering in the $p$-wave shows unambiguously that the $\rho(770)$ and $K^{*}(892)$ have the $N_{c}$ dependence expected of \emph{ordinary} mesons. \begin{figure}[ht]
\includegraphics[width=14cm]{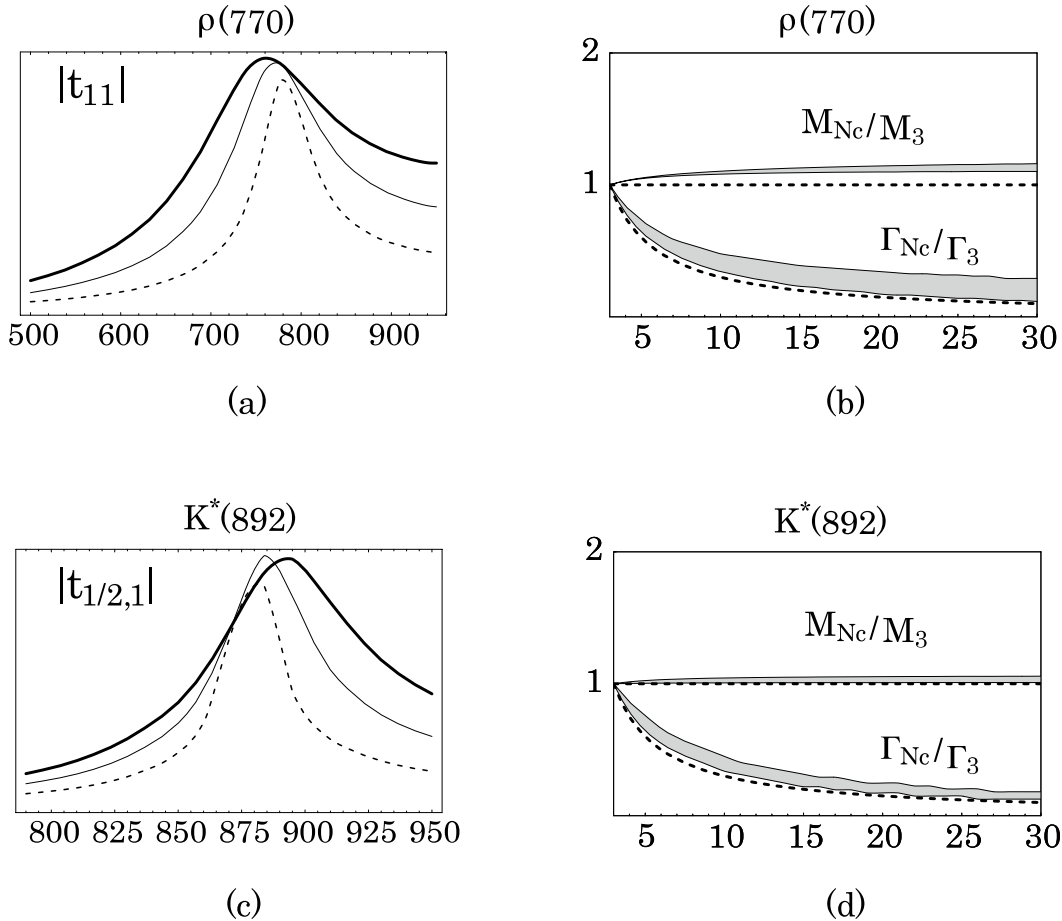}
\caption{\small  $\pi\pi$ ((a)and (b)) and $\pi K$ ((c) and (d)) scattering in the $p$-wave for various values of $N_{c}$ as extrapolated from chiral perturbation theory.  In (a) and (c) the modulus of the scattering amplitude  is plotted versus the center of mass energy (in MeV).  In (b) and (d) the mass and width are plotted relative to their values at $N_{c}=3$.  The dark, solid and dotted curves in (a) and (c) correspond to $N_{c}=$3, 5, and 10 respectively.  The dashed line in (b) and (d) show the expectation from standard $N_{c}$ counting arguments.  From Ref.~\cite{Pelaez:2003dy}.}
\label{pwaves}
\end{figure}
Figs.~\ref{pwaves}(a) and \ref{pwaves}(c) from Ref.~\cite{Pelaez:2003dy} shows the $N_{c}$ dependence of the $p$-wave $\pi\pi$ and $K\pi$ amplitudes.  Figs. \ref{pwaves}(b) and \ref{pwaves}(d)  summarize the $N_{c}$ dependence of the mass and width of the $\rho(770)$ and $K^{*}(892)$ resonances extracted from the curves in (a) and (c). The figures confirm the expectations of a large $N_{c}$ analysis. 
Both the $\rho$ and the $K^{*}$ appears clearly for all $N_{c}$.  Their masses change little as $N_{c}$ increases.  However their widths  decrease  significantly with $N_{c}$.  In fact, they fit well the $1/N_{c}$ dependence expected from the large $N_{c}$ analysis.

 So unitarized chiral perturbation theory gives strong corroboration that the $\rho(770)$ and the $K^{*}(892)$ are \emph{ordinary} hadrons.  The $\pi\pi$ and $\pi K$ $s$-waves behave totally differently, and are discussed in the next section.

\subsection{Ordinary hadrons as Feshbach resonances}
Suppose, for the sake of argument, that the $\rho$ appeared at a low enough energy in $\pi\pi$ scattering   that the use of non-relativistic scattering theory was justified.  How would it be interpreted?  Unlike the relativistic case, non-relativistic scattering theory is completely understood in terms of a multi-channel \sch equation, so we would have to seek a description of the $\rho$ in that language.  The $\rho$ owes its stability to the fact that once a quark in one pion has annihilated with an antiquark in the other, the remaining quark-antiquark pair lie in a confined channel in the Hilbert space, a channel with a discrete spectrum.  If the $\bar qq$ annihilation amplitude vanished, as it does as $N_{c}\to\infty$, the confined channel would decouple from $\pi\pi$, and the $\rho$ would become a \emph{bound state in the $\pi\pi$ continuum}.  This way of generating a resonance has been known since the early days of quantum mechanics\cite{rice}.  Fano applied it to electron atom scattering\cite{fano} and Feshbach employed it with great success in nuclear scattering theory\cite{feshbach}, and the phenomenon has become known as a \emph{Feshbach resonance}.

Although Feshbach resonances are well-known to atomic and nuclear physicists, they are not so familiar to the QCD community.  The basic idea  can be illustrated by studying an intentionally simplified model of non-relativistic scattering of two spinless particles with two channels, one open and the other confined.  Channel 1, the \emph{open channel}, has a continuum beginning by convention at $E=p^{2}=0$, and for simplicity, \emph{no interaction potential at all}.  Channel 2, the \emph{confined channel}, has only a discrete spectrum.  The two channels are coupled by an interaction potential, $V$, which for the sake of analogy to QCD, we take to be of order $1/\sqrt{N_{c}}$.  The generalization to many open and/or many confined channels is straightforward\cite{feshbach}.   The Hamiltonian for this system is given
by
\begin{equation} 
\mathcal{H}=
\begin{pmatrix}
h_{0} & V  \\
V & h
\end{pmatrix}\,.
\label{eq1}
\end{equation}
$h_{0}$ is the Hamiltonian for free motion in the open channel, for example $h_{0}\equiv h_{\ell}=-  \frac{d^{2}}{dr^{2}}+\frac{\llll}{r^{2}}$ in the $\ell^{\rm th}$ partial wave ($\hbar^{2}/2m=1$ for convenience).
The spectrum of $h$, the Hamiltonian restricted to the confined channel, is discrete.  Let $\brak{\phi}$ be its lowest eigenstate --- the ``proto''-$\rho$ --- with energy $E_{0}$, $h\brak{\phi}=E_{0}\brak{\phi}$.
 
At any energy the effect of the coupling to the closed channel can be included in the open-channel \sch equation by introducing the Green's function,
\be
\la{greensfunction}
{\cal G}(E)=\frac{1}{E-h}
\ee
so that
\be
\la{effective}
h_{\ell}\brak{u_{\ell}}+V{\cal G}(E)V\brak{u_{\ell}}=E\brak{u_{\ell}}.
\ee
Note that the \emph{only} interaction in the open channel comes from its coupling to the confined channel.

At energies close to $E_{0}$, the problem simplifies because the closed channel Green's function can be approximated by,
\be
\la{gf}
{\cal G}(E)=\frac{1}{E-h}\approx \frac{\brak{\phi}\krab{\phi}}{E-E_{0}},
\ee
allowing the \sch equation in the open channel to be written in terms of an energy dependent,  separable effective potential, 
\be
\la{eff}
h_{\ell}\brak{u_{\ell}}+V\frac{\brak{\phi}\krab{\phi}}{E-E_{0}}V\brak{u_{\ell}}=E\brak{u_{\ell}}.
\ee
or in coordinate space,  
\be
\la{schr}
-u''_{\ell} (r)+\frac{\ell(\ell+1)}{r^{2}}u_{\ell}(r)+ \frac{1}{E-E_{0}} \int_{0}^{\infty}dr'v(r)v(r')u_{\ell}(r')=Eu_{\ell}(r'),
\ee
where   $ v(r)\equiv\krab{r}V\brak{\phi}=V(r)\phi(r)$.  Here $\phi(r)$ is the radial wavefunction of the state $\brak{\phi}$ and $V(r)$ is the interaction potential in coordinate space\footnote{I have ignored the angular dependence of the potential necessary to effect the  transition from the continuum with orbital angular momentum $\ell$ to the discrete confined state with angular momentum $j$. In the case of the $\rho$, for example, the $\pi\pi$ $p$-wave must couple to the $s$-wave $\bar q q$ spin one bound state.}.

This equation is valid only at energies close enough to $E_{0}$ that the pole $\sim\frac{1}{E-E_{0}}$ dominates ${\cal G}(E)$.
It is elementary to solve for the scattering amplitude in the open channel in terms of the matrix element,
\be
\la{xi}
\xi(E)=\krab{\phi}V\brak{u_{\ell}^{0}(E)}
\ee
where $u_{\ell}^{0}(E)$ is the (continuum normalized) solution to \eq{schr} with $v(r)=0$.  For angular momentum $\ell$ in three dimensions, $\brakket{r}{u_{\ell}^{0}(E)}=r j_{\ell}(pr)$, so
\be
\la{xiform}
\xi_{\ell}(E)=\int_{0}^{\infty} dr rj_{\ell}(pr)v(r), 
\ee
and the Argand amplitude is
\be
\la{ff} 
 f^{F}_{\ell}(E) \equiv e^{i \delta_{\ell}(E)} \sin\delta_{\ell}(E) =
  \frac{\sqrt{E}  \xi^2_{\ell}(E)}{E_{0}-E - 
X_{\ell}(E)}. 
\ee
where
\begin{equation}
\la{denom}
X_{\ell}(E) = \frac{1}{\pi}\int\limits_0^{\infty}dE' 
\frac{\sqrt{E'}}{E'-E-i\epsilon}  \xi^2_{\ell}(E').
\end{equation} 
Eqs.~(\ref{ff}) and (\ref{denom}) summarize the way that Feshbach resonances (hence the superscript ``$F$'' on $f_{\ell}^{F}$) are manifested in a partial wave scattering amplitude.  It is easy to see that ${\rm Im}f_{\ell}^{-1}(E)=-1$ as required by unitarity, so eq.~(\ref{ff}) can be rewritten as 
 \begin{equation}
\la{unitary}
 f_{\ell}(E) =
  \frac{\sqrt{E}  \xi^2_{\ell}(E)}{E_{0}-E - i\sqrt{E}\xi^{2}_{\ell}(E)-\frac{1}{\pi}{\rm P.V.}\int_0^{\infty}dE' 
\frac{\sqrt{E'}}{E'-E}  \xi^2_{\ell}(E')}.
\end{equation} 
where ``P.V.'' denotes the principal value.  As long as $ \xi^{2}_{\ell}(E_{0})\ll \sqrt{E_{0}}$, eq.~(\ref{unitary}) always has a resonance pole below the real axis in the complex $E$ plane and close enough to $E_{0}$ to be well approximated by 
\be
\la{feshres}
E_{\rm res}=E_{0}- ip_{0}\xi^{2}_{\ell}(E_{0})-\frac{1}{\pi}{\rm P.V.}\int_0^{\infty}dE' 
\frac{\sqrt{E'}}{E'-E_{0}}  \xi^2_{\ell}(E')
\ee
(where $E=p^{2}$).
The function $\xi_{\ell}(E)$ controls the width:  the more weakly the open channel   is coupled by the interaction $V$ to the discrete state in the closed channel, the narrower the resonance.

As an illustration, to make the whole discussion more concrete, I have made   a specific choice for the product of the channel coupling potential and the bound state, and computed the phase shift for a Feshbach resonance in  channels with $\ell=0$ and 1.  Most of the calculation can be performed analytically if we choose 
\be
\la{form}
v(r)= \sqrt{\frac{4\mu^{2}\lambda}{N_{c}r}}e^{-\mu r} .
\ee
To mimic QCD I have introduced a factor of  $1/\sqrt{N_{c}}$ into the coupling between open and closed channels.  This leaves $\sqrt{\lambda}$ as a dimensionless   and $N_{c}$ independent measure of the strength of the coupling.  $v(r)$ has been normalized or equivalently, $\lambda$ has been chosen, so that $\int dr r^{2}v^{2}(r)=1$.  $1/\mu$ is the range of the coupling to the open channel, determined in reality by both the size of the discrete state and the character of the transition potential.  The advantages of this choice for $ v(r)$ are first, that it has finite range, as one would expect on physical grounds, and second, that all the calculations can be performed analytically.  The transition from the non-relativistic kinematics of the original \sch equation, where $E=p^{2}$, to the relativistic case, where $E=2\sqrt{p^{2}+m^{2}}$ is accomplished by switching to $p$ as the independent variable.  Then the scattering amplitude   will have   the proper analytic behavior as a function of the center of mass momentum $p$.   
  
\begin{figure}[ht]
\includegraphics[width=14cm]{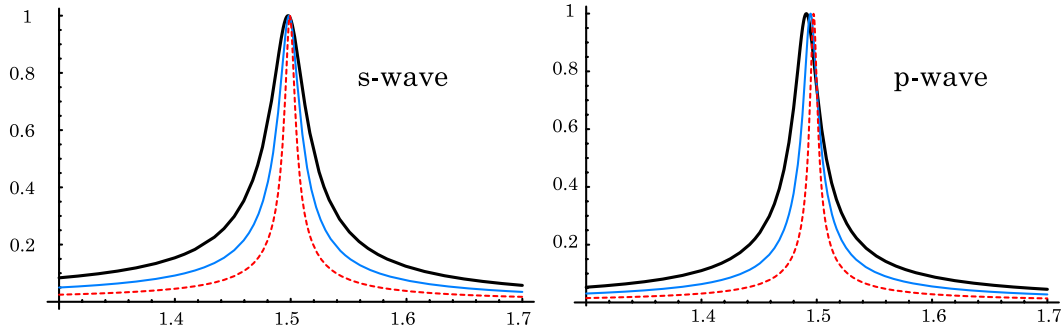}
\caption{\small (a) $s$-wave and  (b) $p$-wave Feshbach resonances.  The absolute value of $f_{\ell}(E)\equiv \sin\delta_{\ell}(E)$ is plotted versus $p=\sqrt{E}$.  The three curves correspond to scaling the coupling as if $N_{c}\to$ 3, 5, and 10.  The bound state in the continuum was somewhat arbitrarily set at $E_{0}=2.25\mu^{2}$.}
\label{feshbach}
\end{figure}

The $s$-wave and $p$-wave phase shifts obtained from this model are shown in Fig.~\ref{feshbach} for a fixed $\lambda$ and  $N_{c}=$ 3, 5, and 10.  In  both cases the width of the resonance goes to zero and its location approaches a fixed value as $N_{c}\to\infty$.  Of course, this behavior is guaranteed by the structure that was put into the model:  a confined state in the continuum.  It would be quite straightforward to describe any narrow meson resonance in two body scattering in this manner.

Note in particular that  the Feshbach mechanism generates a resonance in the $s$-wave is as naturally as in any other partial wave.  In contrast it is quite unnatural to generate an $s$-wave resonance in single channel potential scattering.  A resonance occurs when an attraction is counter balanced by a repulsive barrier.  For higher partial waves, angular momentum provides the barrier, but in the $s$-wave, the potential itself would have to change from attractive at short distances to repulsive at longer distances. A  Feshbach resonance, on the other hand, requires only that the open channel, whatever its angular momentum, couples to a confined channel.  It is interesting, therefore, that narrow $s$-wave resonances are not known in QCD.   Instead, as recently summarized by Rosner \cite{Rosner:2006vc}, bound states, virtual states, and broad enhancements are the common features found in the $s$-wave, which leads us to consider extraordinary mesons.  I return to this issue in the last section of this talk.

\section{Extraordinary Mesons}

\subsection{Large $N_{c}$ expectations}

What can be said of meson-meson scattering at large $N_{c}$ at energies away from the narrow resonances discussed in the previous section?  The qualitative predictions can be found by a minimal extension of 't Hooft's large $N_{c}$ analysis  to this case\cite{Jaffe:1981up}.  First we need a properly normalized interpolating operator that creates two color-singlet mesons,
\be
D(x)\equiv \frac{1}{N_{c}} \bar q q \bar q q(x)\, . \la{2.1}
\ee
The fact that I have chosen a local $\bar q \bar q q q$ operator rather than the product of two $\bar qq$ operators at different spacetime points makes no difference in the classification of physical processes at large $N_{c}$.  Questions like ``do stable or resonant $\bar q \bar qqq$ states exist?'' depend on the dynamics, not the choice of interpolation field.  The internal color coupling does not matter either, because any $\bar q\bar q qq $ operator which is an overall color singlet can be Fierz-transformed into color singlet quark bilinears:
\be
\la{2.2}
D(x)=\cos\theta\  M_{12}(x)M_{34}(x)+\sin\theta\ 
M_{14}(x)M_{23}(x)
\ee
where $M_{ij}(x)=[\bar q_{i}(x)q_{j}(x)]^{\bf 1}$ is a color singlet (meson) operator, and the indices, $i,j,...$ refer to all labels other than color ({\it eg.\/} spin, flavor, space).  The factor of $N_{c}$ in \eq{2.1} was chosen so that the correlation function, 
$\Pi(x)= \krab{0}T(D(x)D(0))\brak{0}$ is normalized to unity as $N_{c}\to\infty$.  

 \begin{figure} 
\includegraphics[width=14cm]{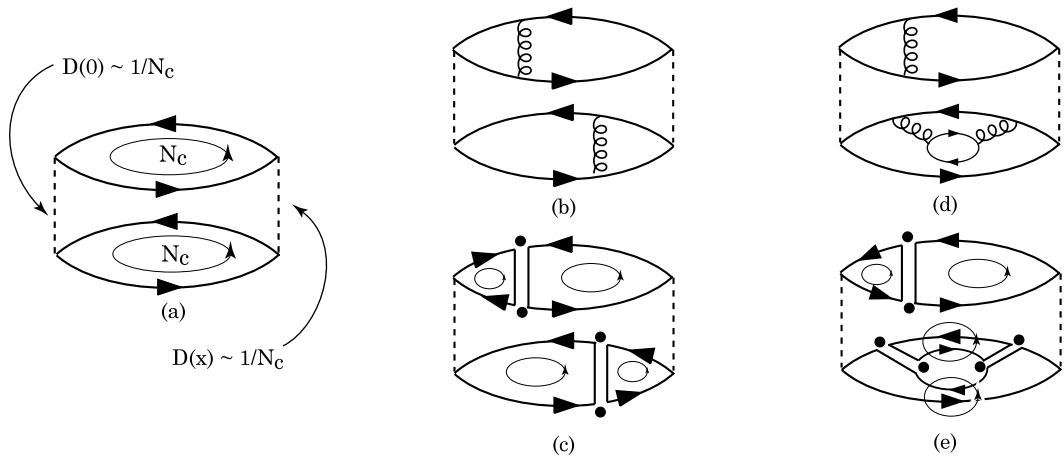}
\caption{\small Feynman and 't Hooft diagrams for $\bar q \bar qqq$ sources.  Each   loop carries a factor of $N_{c}$, each vertex (denoted with a dot), $1/\sqrt{N_{c}}$, and each factor of $D(x)$ carries $1/N_{c}$.  For clarity  each source  is pulled apart into separate $\bar q q$ pairs connected by a dotted line.  (a) Normalization of the $\bar q \bar q q q$ source, $D(x)$. (b) and (c) typical Feynman and 't Hooft diagrams for QCD interactions within color singlet $\bar qq$ pairs, which are ${\cal O}(1)$ at large $N_{c}$.  (d) and (e) quark pair creation within a $\bar q \bar qqq$ correlator is suppressed by $1/N_{c}$.  }
\label{mma}
\end{figure}
\begin{figure}[ht]
\includegraphics[width=14cm]{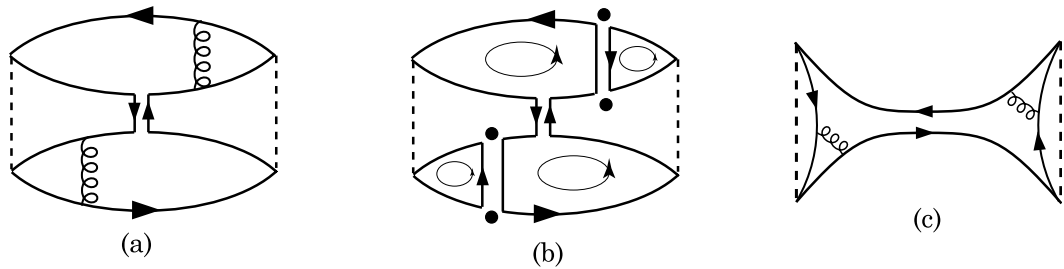}
\caption{\small $t$ channel meson exchange is suppressed by $1/N_{c}$:  (a) Generic Feynman diagram; (b) 't Hooft notation; (c) Feynman diagram redrawn to emphasize duality with $s$-channel (Feshbach) resonance formation.}
\label{mmd}
\end{figure}

\begin{figure}[ht]
\includegraphics[width=10cm]{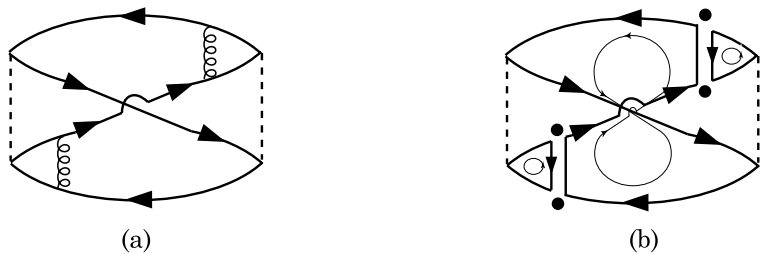}
\caption{\small Quark exchange between color singlet mesons contributes at ${\cal O}(1/N_{c})$ to meson-meson scattering away from the narrow resonances:  (a) Generic Feynman diagram; (b) 't Hooft notation.}
\label{mme}
\end{figure}

\begin{figure}[ht]
\includegraphics[width=10cm]{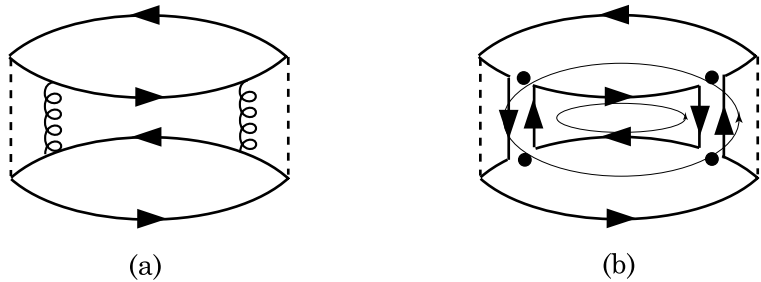}
\caption{\small Exchange of vacuum quantum numbers between mesons is down by $(1/N_{c}^{2})$ as $N_{c}\to\infty$.  (a) Generic Feynman diagram; (b) 't Hooft notation.}
\label{mmf}
\end{figure}

  
  The Feynman graphs contributing to $\Pi(x)$ fall into classes that characterize meson-meson scattering.  The important categories are shown in Fig.~\ref{mma}--\ref{mmf}.  [For easier visualization, the $D(x)$ vertex is drawn with the two $\bar q q$ pairs separated by a dotted line.]  
The leading contributions, shown in Fig.~\ref{mma}(b) and (c)  are \emph{disconnected diagrams} from the point of view of meson-meson scattering.   From Fig.~\ref{mma} (a)  it is clear that $D$ is normalized to unity, and Fig.~\ref{mma} (b) and (c)  show  that interactions within color singlet mesons persist at large $N_{c}$.  These are the only diagrams that survive   as $N_{c}\to\infty$.  Diagrams like Fig.~\ref{mma} (d) and (e) that mix quark-antiquark pairs into the meson wavefunctions are ${\cal O}(1/N_{c})$, so quark (and antiquark) numbers are conserved in this limit.

What about interactions between mesons?  Exchange of ordinary mesons in the $t$-channel, shown in Fig.~\ref{mmd}, contributes at ${\cal O}(1/N_{c})$ (there is one fewer loop  for the same number of vertices compared to Fig.~\ref{mma} (c)).  However duality, which is expected to become exact as $N_{c}\to\infty$, identifies the sum over all $t$-channel exchanges of the form of Fig.~\ref{mmd} (a) with the tower of ordinary $s$-channel resonances that have widths of order $1/N_{c}$.  The $N_{c}$ counting works out because the effect of the $s$-channel resonances on the scattering amplitude averaged over energy is proportional to their width, ${\cal O}(1/N_{c})$.  We conclude that the ${\cal O}(1/N_{c})$ effects of $t$-channel exchange of ordinary mesons is manifested in $s$-channel resonances, so any possible effect on the meson-meson continuum vanishes \emph{faster than $1/N_{c}$}.  So far the meson-meson continuum is unaffected by interactions at order $1/N_{c}$.  

The only remaining contribution to off-resonance meson-meson scattering of order $1/N_{c}$  comes from quark exchange diagrams, shown in Fig.~\ref{mme}.  It is easy to see that 
these diagrams go like $1/N_{c}$ as $N_{c}\to\infty$.  They are associated with terms of the form,
\be
\la{2.3}
\Pi(x)\sim \krab{0}T\left(\left[M_{12}(x)M_{34}(x)\right]\left[M_{14}(0)M_{32}(0)\right]\right)\brak{0}
\ee
in the notation of eq.~(\ref{2.2}).  Note the exchange of spin, flavor and/or space indices.  Although the color structure of the wavefunction is scrambled by quark exchange, the resulting configuration can always be written in terms of pairs of color singlet mesons.  Therefore these diagrams generate forces in the space of color singlet mesons and should be described by some relativistic generalization of a potential in the space of meson-meson scattering states.  They do not couple the scattering mesons to the narrow $\bar qq$ meson resonances.

Contributions to meson-meson scattering that do not involve quark exchange, see for example  Fig.~\ref{mmf}, vanish like $1/N_{c}^{2}$ as $N_{c}\to\infty$.  It is easy to see that production of other mesons in the final state is further suppressed by a factor of $1/\sqrt{N_{c}}$ for each additional meson.  So we conclude that through order $1/N_{c}$, quark exchange is the sole contribution to meson-meson  scattering off resonance. \emph{It dominates as $N_{c}\to\infty$.}

The observation that meson-meson scattering is dominated by the quark exchange diagram of Fig.~\ref{mme} off resonance leads to the following conclusions (at order $1/N_{c}$):  
\begin{itemize}
\item In any \emph{fixed} basis, {\it eg.\/} $M_{12}M_{34}$, the exchange mixes color octet components into the $(\bar q q)(\bar q q)$ wavefunction.  So the origin of the interaction is chromodynamic rather than hadronic.  It should not be described by an effective Lagrangian in which only meson degrees of freedom appear.
\item The range of the interaction is determined by the distances at which the quarks in the scattering mesons overlap, {\it i.\,e.\/} a typical confinement scale of order 1 Fermi, not by the masses of exchanged mesons\footnote{The fact that the range of the interaction is determined by the ``size'' of the hadrons and not by the mass of the lightest exchanged meson is discussed in Refs.~\cite{Jaffe:1990xj}.  It is due to dominance of the ``would-be'' anomalous threshold in QCD.}.
Since these interactions are finite range they are likely to be most important in low angular momentum partial waves, $\ell=0$ in particular.  The interaction might be attractive or repulsive.  

\smallskip
\noindent
There are other reasons to think that the interaction is repulsive in exotic $\bar q\bar q qq$ channels, accounting for the absence of exotics\cite{Jaffe:1976ig,Jaffe:2004ph}, and attractive in channels dominated by the color, spin and flavor antisymmetric diquark.   If attractive  it can generate bound or virtual states or enhancements in the $s$-wave and/or bound states or resonances in higher partial waves.  These are the ``extraordinary'' hadrons. 
\item The interaction of Fig.~\ref{mme} does not couple the meson-meson system to confined channels.  It must be described entirely in terms of the scattering states of the meson-meson system.  If the problem were non-relativistic, the natural framework would be the \sch equation in the channel space of mesons, $\pi\pi$, $K\overline K$, $\eta\eta$, {\it etc.\/} ({\it e.\,g.\/} for $I=J=0$).  The generalization of non-relativistic potential scattering to relativistic circumstances has never been clear, and remains problematic in this case.  Pions are already relativistic in the energy region of the $f_{0}(600)$, so the \sch equation won't suffice for a quantitative study of this problem.
\item Extraordinary hadrons disappear as $N_{c}\to\infty$.  The interaction that produces them goes away and the meson-meson scattering state turns into the non-interacting continuum.
\end{itemize}
Extraordinary hadrons could not be more different from ordinary hadrons.  Of course they are only interesting if they exist in QCD.  Study of chiral dynamics at large $N_{c}$ suggests that the $f_{0}(600)$ is an extraordinary hadron.

\subsection{Evidence from chiral dynamics}

Pel\'aez has applied the techniques described in Section \ref{chiralpwave} to the $\pi\pi$ and $\pi K$ $s$-wave scattering amplitudes at low energy, where both channels show broad enhancements.  The $\pi\pi$ enhancement, the $f_{0}(600)$, is now a recognized ``resonance''.  The $\pi K$ enhancement, known as the $\kappa(800)$, is well documented, but the existence of an $S$-matrix pole (and therefore the use of the term ``resonance'') is controversial.  These results differ dramatically from the $p$-waves that were shown in Fig.~\ref{pwaves}.
\begin{figure}[ht]
\includegraphics[width=14cm]{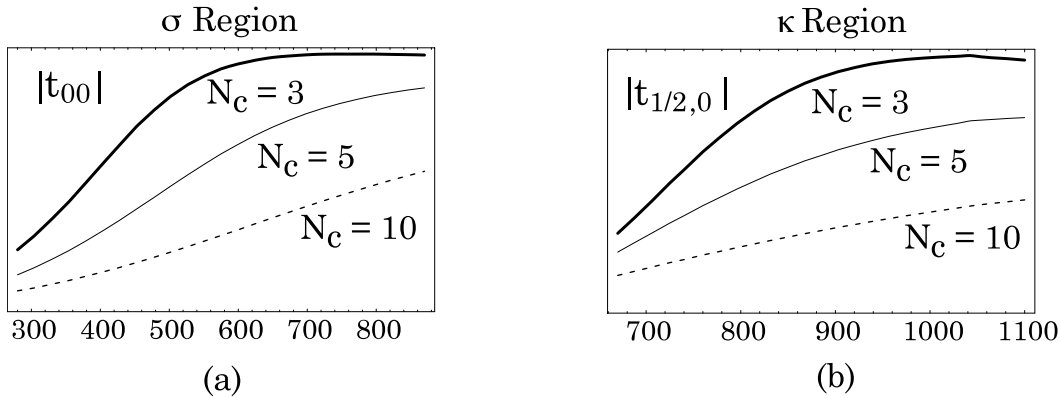}
\caption{\small (a) $\pi\pi$ and (b) $\pi K$ scattering in the $s$-wave for various values of $N_{c}$ as extrapolated from chiral perturbation theory.  The modulus of the scattering amplitude, $|\sin\delta|$ is plotted versus the center of mass energy (in MeV).  From Ref.~\cite{Pelaez:2003dy}.}
\label{swaves}
\end{figure}  
As shown in Fig.~\ref{swaves}, the broad enhancements in both channels apparent at $N_{c}=3$ increase  in width with increasing $N_{c}$ and subside  into the continuum at the largest values of $N_{c}$.  This is precisely the expected behavior for   extraordinary hadrons.  

\begin{figure}[ht]
\includegraphics[width=14cm]{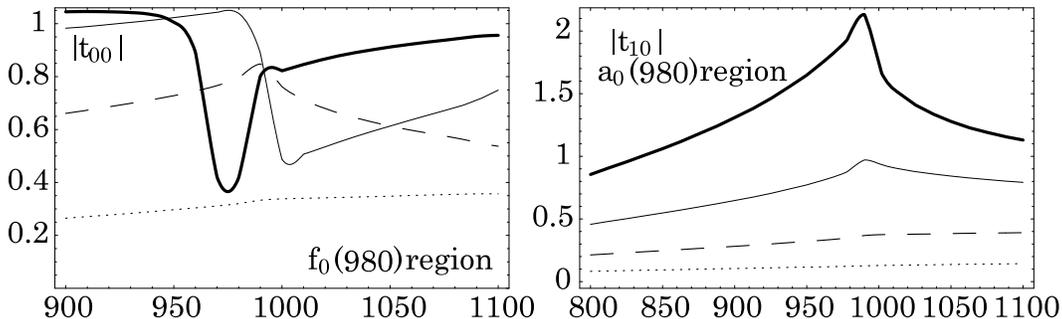}
\caption{\small Moduli of the $I=J=0$ and $I=1$, $J=0$ meson-meson amplitudes are plotted versus center of mass energy for $N_{c} = 3$ (thick line) $N_{c} = 5$ (thin continuous line), $N_{c} = 10$ (dashed) and $N_{c} = 25$ (thin dotted line).  The complex structures near $K\overline K$ threshold, known as the $f_{0}(980)$ and $a_{0}(980)$ disappear as $N_{c}$ increases.  See the discussion in Ref.\cite{Pelaez:2003dy} concerning the robustness of this result.}
\label{kkbar}
\end{figure}

Pel\'aez extended his analysis of pseudoscalar meson-meson scattering all the way to $K\overline K$ threshold, quite an extrapolation for chiral perturbation theory.  With that caveat and the warning that the IAM is not unique in a multichannel environment, his results suggest that the $f_{0}(980)$ (coupling to $\pi\pi$ and $K\overline K$) and, less certainly, the $a_{0}(980)$ (coupling to $\pi\eta$ and $K\overline K$) also behave like extraordinary hadrons.  Fig.~(\ref{kkbar}) shows his results for these channels.  In the case of the $a_{0}(980)$ there are regions at the margin of Pel\'aez's parameter space where the state persists at large $N_{c}$, so the interpretation of this state as an extraordinary hadron is less certain.

So far these are the only reliable identifications of observed effects that may be examples of a different class of hadrons.  It is interesting that the  $f_{0}(980)$, $a_{0}(980)$, $\kappa(800)$, and $f_{0}(600)$ fill up an $SU(3)_{\rm f}$ nonet, suggesting that they may have similar origins.  These objects are discussed further in the last section.

\subsection{A toy model}
 
There is no satisfactory way to generalize potential scattering to relativistic systems.  Not only is there no natural generalization of the \sch equation to relativistic kinematics, but also the constraint of  unitarity in crossed channels, absent in the non-relativistic limit, makes it unclear how to translate a physical model for interactions in the $s$-channel into an fully unitary scattering amplitude.  

Nevertheless, it seems useful to present  some, albeit crude, model of how a potential-like interaction might generate enhancements, bound states, resonances and the like in a relativistic scattering amplitude.  The $N/D$ method of Chew and Mandelstam\cite{chewmandelstam} is a general, relativistic approach that respects two-body unitarity in the $s$-channel and mimics potential scattering.   They developed the method in a very ambitious attempt to implement all the constraints of unitarity and analyticity.  I apologize for recycling it in such a humble application.  
\begin{figure}[ht]
\includegraphics[width=14cm]{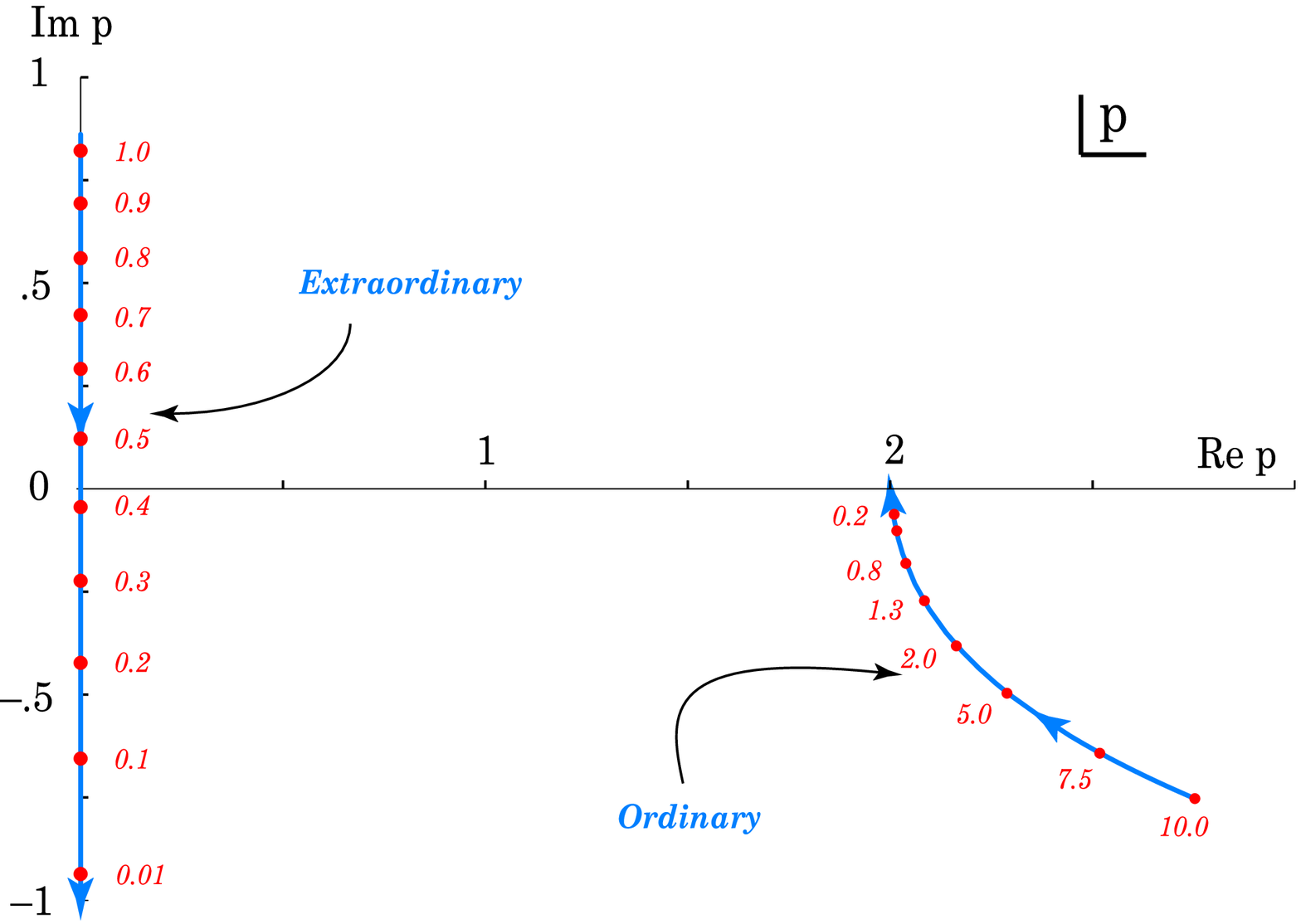}
\caption{\small 
  The trajectories of poles in the complex $p$-plane are shown for the  
$s$-wave scattering amplitudes in two toy models.  The trajectory labeled ``ordinary'' comes from the Feshbach resonance model of \eq{ffagain}.  The trajectory labeled ``extraordinary'' comes from the $N/D$ model embodied in \eq{ndampl}.  The point on the 
trajectories are labeled by values of $\lambda/N_{c}$.  A pole on the positive imaginary axis is a bound state; one on the negative imaginary axis near the origin is a virtual state; and one just below the real $p$-axis is a resonance.  Note the opposite behavior of the two types of states.
 } 
\label{compares}
\end{figure}
\begin{figure}[ht]
\includegraphics[width=14cm]{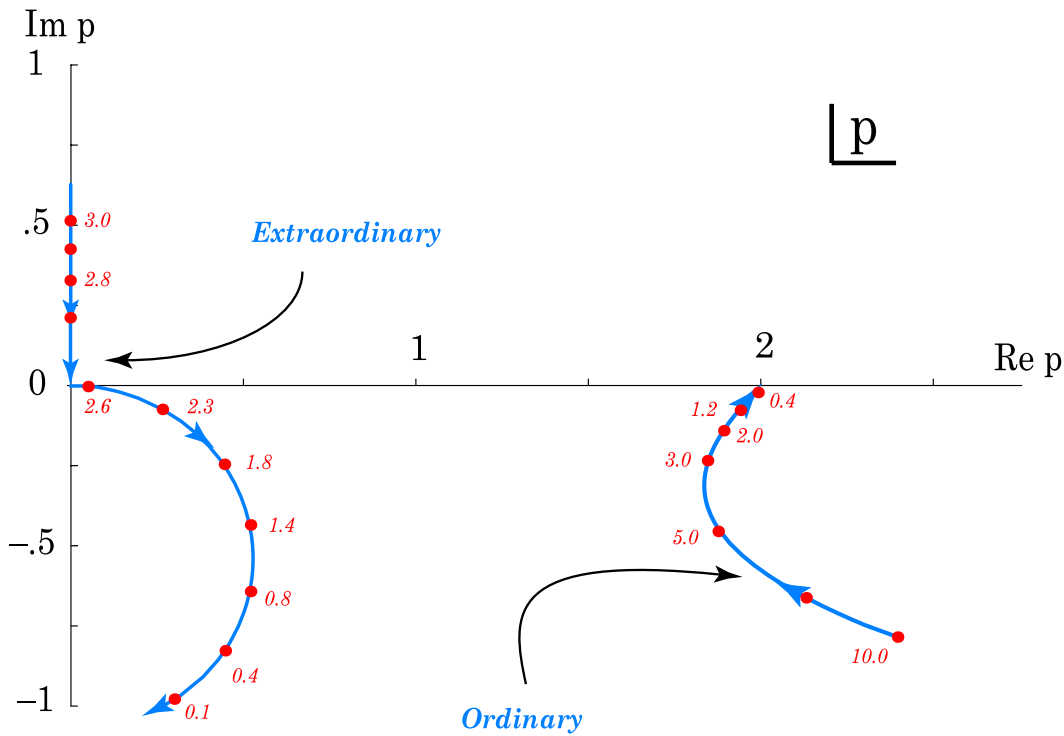}
\caption{\small  The trajectories of poles in the complex $p$-plane are shown for the  
$p$-wave scattering amplitudes in two toy models.  The trajectory labeled ``ordinary'' comes from the Feshbach resonance model of \eq{ffagain}.  The trajectory labeled ``extraordinary'' comes from the $N/D$ model embodied in \eq{ndampl}.  Points on the 
trajectories are labeled by values of $\lambda/N_{c}$.  A pole on the positive imaginary axis is a bound state  and one just below the real $p$-axis is a resonance.  Note the opposite behavior of the two types of states.}
\label{comparep}
\end{figure}
Let $f_{IJ}(p)$ be the $\pi\pi$ Argand amplitude ($f_{IJ}=e^{i\delta_{IJ}}\sin\delta_{IJ}$) with isospin $I$ and angular momentum $J$, as a function of the pion momentum in the center of mass.  Then a form of $f_{IJ}$ consistent with $s$-channel unitarity is
\be
\la{noverd}
f_{IJ}^{N/D}(p)\equiv \frac{pN(p)}{D(p)}= \frac{pN(p)}{1-\frac{2}{\pi}\int_{0}^{\infty}dq q^{2}\frac{N(q)}{q^{2}-k^{2}-i\epsilon}}
\ee  
provided that $N(p)$ is a real analytic function of $p^{2}$ that has singularities only on the negative $p^{2}$ axis (left-hand singularities). Then the denominator function, $D(p)$, has no left hand singularities and has the required unitarity cut (a ``right hand singularity'') beginning at $p=0$, specifically ${\rm Im} (f_{IJ}^{N/D})^{-1}(p)=-1$ for $p>0$, as can easily be verified from \eq{noverd}.  

  A simple introduction to the $N/D$ method and its connection both to the Born expansion and to the effective range approximation can be found in Ref.~\cite{frautschi}.\footnote{My parameterization of $D(p)$ differs from Frautschi's by the ``subtraction constant'', so that $\lim_{p\to\infty} D(p)=1$.  This choice emerges from the separable potential example (see below).}   Imagining $pN(p)\ll 1$ and expanding the denominator allows one to make contact with the Born expansion, where  one sees that $N(p)$ is proportional to the Fourier transform of the 
$s$-channel potential.  Thus $pN(p)/D(p)$ represents a unitarization of the first Born approximation in the same way that the scattering amplitude that comes from solving the \sch equation with a potential $V(\vec x)$ represents a unitarization of the first Born approximation, $\widetilde V(\vec q)$.  The $pN(p)/D(p)$ form is not unique --- for example CDD poles can be added to $D(p)$.  

If the input $N(p)$ is large enough, $D(p)$ can vanish, giving rise to a pole near the physical region, which can appear as a resonance, virtual state or bound state, depending on the angular momentum and the strength and form of $N(p)$.  The dependence of the pole location, and therefore of its physical effects, on the strength of the interaction are dramatically different from the behavior of a Feshbach resonance.

To make this discussion more concrete and to make a comparison with the Feshbach resonances of the previous section easy, I consider  we consider a specific model for $N(p)$, namely the same function $\xi_{\ell}^{2}(E)$ that appeared in the analysis of Feshbach resonances.
\be
\la{ndampl} 
f^{N/D}_{\ell}(E)=\frac{\lambda\frac{\sqrt{E}}{N_c} \overline\xi_{\ell}^{2}(E) }{1-\frac{2\lambda}{N_c\pi}\int_{E_{\rm th}}^\infty
dE'\sqrt{E'}\frac{\overline\xi_{\ell}^{2}(E') }{E'-E-i\epsilon}}
\ee
For easy comparison with the Feshbach resonance amplitude I have switched back to $E$ as the independent variable.  Also for clarity I redefined $\xi_{\ell}(E)$ to make the dependence on the coupling strength, $\lambda$, and $N_{c}$ explicit:  $\xi_{\ell}^{2}(E)  =\frac{\lambda}{N_{c}}\overline\xi_{\ell}^{2}(E)$.\footnote{I have also suppressed the labels $I$ and $J$, but preserved $\ell$ which labels the meson-meson partial wave.}  
The scattering amplitude in the case of a Feshbach resonance can be  written in the same notation for comparison (see \eq{ff}), 
\be
\la{ffagain} 
f^{F}_{\ell}(E)=\frac{\lambda\frac{\sqrt{E}}{N_c}\frac{1}{E-E_{0}}\overline\xi_{\ell}^{2}(E)}{1-\frac{1}{E-E_{0}}\frac{2\lambda}{N_c\pi}\int_{E_{\rm th}}^\infty
dE'\sqrt{E'}\frac{\overline\xi_{\ell}^{2}(E')}{E'-E-i\epsilon}}
\ee
The only, but crucial, difference is the pole in $N(E)$ in the Feshbach amplitude, $f^{F}_{\ell}$, generated by the propagation in the confined channel.

These two models are even more closely related.  The $N/D$ scattering amplitude is the solution to the \sch equation with a separable potential,
\be
\la{noverd2}
h_{\ell}\brak{u_{\ell}}+V\brak{\phi}\krab{\phi}V\brak{u_{\ell}}=E\brak{u_{\ell}},
\ee
which is identical to \eq{eff} except for the singular energy dependence introduced by the factor of $1/(E-E_{0})$ in \eq{eff}.

The difference between ordinary and extraordinary hadrons can be brought into sharp focus if we compare the structure of the $S$-matrix in these two toy models.  As discussed in the previous section, $f^{F}_{\ell}$ has a pole at $E\approx E_{0}$ when $N_{c}$ is large (or $\lambda$ is small).  The $s$-wave and higher partial waves do not differ dramatically in the Feshbach resonance model.  In contrast, $f^{N/D}_{\ell}$ can develop a pole near the physical region only when $\lambda$ is large.  In the $s$-wave that pole appears on the negative imaginary $p$-axis and moves toward $p=0$ as $\lambda/N_{c}$ \emph{increases}.  It is manifested in the scattering first as a virtual state, when ${\rm Im}\ p\lesssim 0$, and then as a bound state, when ${\rm Im} \ p>0$.  In contrast, in higher partial waves, the pole approaches the physical region as it moves in the fourth quadrant of the complex-$p$ plane --- {\it i.e.\/} as a resonance  whose width decreases as it approaches threshold.  For large enough $\lambda/N_{c}$ is also becomes a bound state. 

The dramatically different dependence of ordinary and extraordinary hadrons on $1/N_{c}$ (at least in these toy models) is illustrated in Figs.~(\ref{compares}) and (\ref{comparep}).  There I have plotted the trajectories of poles in the complex $p$-plane as a function of $\lambda/N_{c}$ for the $s$-wave (Fig.~(\ref{compares})) and $p$-wave (Fig.~(\ref{comparep})) for the simple choice of $\xi_{\ell}(E)$ given in eq.~(\ref{form}).   Note that the singularities move in \emph{opposite directions} as a function of the interaction strength.  Ordinary hadron poles move toward the physical region as the strength diminishes; extraordinary hadrons move away.  The distinction is clear: after all, the models were constructed to make it so.

\goodbreak
\section{Some Thoughts about Extraordinary Mesons}

Having introduced the idea of ordinary and extraordinary mesons, and perhaps convinced the reader that the distinction is interesting, this is a good place to collect some observations that relate to extraordinary mesons and the controversies that surround them.

\subsection{\bf The low energy meson-meson $s$-wave is very special}
It is not at all surprising that the only solid evidence for extraordinary mesons is found in the meson-meson $s$-wave.  Elementary quark model arguments suggest that this is the best channel  in which to see enhancements associated with extraordinary mesons.  First, no ordinary hadronic resonances are expected at low energy in the meson-meson $s$-wave:  to form $0^{+}$ quantum numbers the $\bar qq$ system must be in a \emph{relative} $p$-wave which requires considerable kinetic energy.  All the other $p$-wave $\bar qq$ resonances are well above 1 GeV.  In contrast, the \emph{$s$-wave} $\bar q \bar q qq$ system has $0^{+}$ quantum numbers.  Elementary arguments also suggest that the residual ${\cal O}(1/N_{c})$ interactions that generate extraordinary hadrons are particularly strong in the (scalar diquark)-(scalar antidiquark) channel, which   has $J^{P}=0^{+}$. Contrast this with the meson-meson $p$-wave, with $J^{P}=1^{-}$ quantum numbers, where the reverse is true.  The ordinary resonances like the $\rho(770)$ appear in the $\bar qq$ $s$-wave, while an extraordinary object with minimal quark content $\bar q\bar q qq$ must have a unit of internal excitation to construct negative parity.  
  
  \subsection{\bf Multiquark ``eigenstates'' in a universal confining bag or potential are, in general, not hadronic resonances}
  \label{pmatrix}

A long time ago I calculated the eigenstates of a $\bar q\bar q q q$ system confined to a spherical bag \cite{Jaffe:1976ig}.
The spectrum is discrete because the confining bag boundary condition was applied universally, no matter what the color structure of the $\bar q\bar q qq$ state:  The states do not decay because they are locked in a bag.  The spectrum includes ``states'' with exotic quantum numbers like $\pi^{+}\pi^{+}$.   Does this mean that simple quark models predict lots of exotic meson resonances?  Not at all.  QCD interactions make the exotic ``states'' systematically heavier than $\bar q \bar q qq $ states with non-exotic quantum numbers.  Typically the exotics came out far above threshold for falling apart into pairs of pseudoscalar mesons.  At the time I remarked they were likely just artifacts of imposing confining boundary conditions on channels that are not confined.  Non-exotic states with the quantum numbers of the $f_{0}(600)$, $f_{0}(980)$, $a_{0}(980)$ and $\kappa(800)$, had much lower masses --- a consequence of what would now be called di-quark correlations --- and it was tempting to identify the known enhancements  as $\bar q \bar q qq$ ``states''.  Nevertheless the fact that confining boundary conditions were applied inappropriately in open channels made the identification suspect.

Soon after, Francis Low and I addressed this problem \cite{Jaffe:1978bu} employing a variation of Wigner and Eisenbud's $R$-matrix method\cite{wignereisenbud} for relating discrete spectra computed with convenient boundary conditions to scattering data.   
We found that  quark model eigenstates computed with confining boundary conditions should not automatically be interpreted as evidence for $S$-matrix poles, but instead could be viewed as estimates of \emph{forces} between scattering mesons that are manifested as attractive or even repulsive phase shifts.  If the forces are attractive enough, they could give rise to resonances or bound states.  

Our method and the $R$-matrix method of Ref.~\cite{wignereisenbud} are examples of a general approach to two body, quasi-non-relativistic (both shortcomings) scattering where space is divided in two:  inside some radius $b$ particles interact strongly, outside $b$, weakly or not at all.  In reality the two regions are connected by the continuity of the \sch wave function, but in practice the transition from inside to outside may be hard to compute.  Instead, Wigner suggested, one can solve the inside problem, $H\Psi_{k}=E_{k}\Psi_{k}$ subject to a boundary condition like $\Psi_{k}'(b)=0$ (Wigner's $R$-matrix) or $\Psi_{k}(b)=0$ (our $P$-matrix). That determines the energies at which $R(b,E)\equiv\Psi(b)/\Psi'(b)$ (in Wigner's case) or $P(b,E)\equiv\Psi'(b)/\Psi(b)$ (in our case) has poles as a function of $E$. By continuity, the outside wavefunction must obey the same boundary condition at those energies.   Simple algebraic methods allow one to turn that information into information about the scattering matrix at energies near those poles. 
 
Identifying confining boundary conditions on quarks with a zero in the meson-meson wavefuntion,
Low and I argued that quark model eigenstates might be (at least qualitatively) understood as poles in $P(b,E)$ for a value  of $b$ around the confining radius of $\sim$1 fermi.   The fact that gluon exchange interactions characteristically lower the mass of non-exotic $0^{+}$ $\bar q\bar qqq$ configurations and raise the mass of exotic $0^{+}$ $\bar q\bar q q q $ configurations turned into semi-quantitative information about the phase shifts in meson-meson scattering:   $s$-wave phase shifts
in non-exotic channels should be positive (attractive) and in exotic channels should be negative (repulsive).  As we expected, the existence of quark model eigenstates with exotic quantum numbers \emph{does not lead to exotic resonances}, quite the contrary, it suggests a repulsive interaction in exotic channels.
 
A particularly interest case occurs when, for dynamical reasons, the mass of the quark model $\bar q\bar q q q$ state is below the lightest decay threshold.  It must then correspond to a bound state.  This may occur in Nature:  the lightest $\bar s s (\bar u u+\bar d d)$ and $\bar s s\bar u d$ ``states'' in quark models appear close to $\bar KK$ threshold.  If below threshold, the former can only decay by $\bar s s\to \bar uu$ or $\bar dd$ and then to $\pi\pi$.  The latter still has an allowed decay to $\eta \pi$, although if the strange quark content of the $\eta$ is small, that decay could also be suppressed.  On the basis of this I suggested that the $f_{0}(980)$ and $a_{0}(980)$ might  be $\bar q\bar q qq$ objects in the same $SU(3)_{f}$ multiplet with the \emph{much wider} $f_{0}(600)$ and $\kappa(800)$.  In the language of the present paper, the $f_{0}(980)$ would be a Feshbach resonance, a bound state in $\bar KK$ that couples to the $\pi\pi$ continuum, though the interpretation is complicated by the fact that the $f_{0}(980)$ is so close to $2m_{K}$ that its width overlaps the threshold.
 
 \subsection{\bf There is no clear distinction between a meson-meson molecule and a $\bar q \bar q qq$ state}

The question whether the $f_{0}(980)$ (or the $X(3872)$) is a $\bar q\bar q q q$ state or a $\bar KK$ ($\overline DD^{*}$) molecule could be debated forever.  It is worth remembering, however, that the two cannot be meaningfully distinguished when the state is near or above important thresholds.  The point can be made clear  by considering the deuteron in a world only a little different from ours.  Imagine (1) that the only  force between a proton and neutron came from color exchange at distances less than 1 fermi.  Outside of a range $\lambda\approx 1 \ {\rm fm}$ the proton and neutron don't interact, and (2) that the color exchange force were strong enough to bind the deuteron, but the binding energy is very, very small.  Is this a six quark state or a proton-neutron ``molecule''?  Consider the system as a function of its binding energy, $B$.  When $B$ is small, or more precisely, when $R=\sqrt{\hbar^{2}/2MB}\gg \lambda$, quantum mechanics alone, independent of the nature of the force, dictates that the rms radius of the bound state becomes very large.  The bound state has a probability arbitrarily close to unity to be observed as a proton and neutron separated by an arbitrarily large distance.  However the bound state owes its existence entirely to a six-quark component in its wavefunction that cannot be discovered by considering long range forces between proton and neutron.  It is quite clear what this state \emph{is}.  However calling it a molecule belies its crucial dependence on chromodynamic forces, while calling it a six quark state ignores the dominant term in the wavefunction.   

This reasoning applies to the $f_{0}(980)$ which is very close to $\bar KK$ threshold, and undoubtedly spends most of its time as a separated kaon and anti-kaon.  It also applies to the $X(3872)$ which has similar behavior with respect to $D^{*}\overline D$.

The properties of systems bound by color exchange forces, including many interesting effects on scattering, were explored in a non-relativistic, but instructive toy model by Lenz {\it et al.\/} some years ago \cite{Lenz:1985jk}.

\subsection {\bf Except for very narrow resonances there is no reason to identify $K$-matrix poles with quark eigenstates}

Ideally one would like to have a unitary parametization of the $S$-matrix for  two body scattering in terms of parameters that could be fitted to scattering data and then identified with the QCD eigenstates obtained in models or lattice calculations.  A single, isolated, relatively narrow resonance can be parameterized with a Breit-Wigner shape, and the energy, width, and channel couplings can be extracted from data.  Because the state is quasi-stationary, its properties can in turn be estimated in models or computed using lattice methods that ignore the coupling to decay channels.   Even such a simple case as a resonance over a smooth background or two relatively narrow, but overlapping resonances cannot be treated so simply, because adding ingredients in the $S$-matrix violates unitarity. Long ago, the difficulty of maintaining unitarity when poles are added to one another or to background, motivated a search for a simple, unitary parameterization of the $S$-matrix.  In their analysis of the $\overline KN/\pi \Lambda$ system that led to the discovery of the $\Lambda(1405)$, Dalitz and Tuan \cite{Dalitz:1959dn}\footnote{I believe Dalitz and Tuan were the first to use  the $K$-matrix in  the analysis of hadron-hadron scattering.} proposed to use Pauli's $K$-matrix, which is guaranteed to yield a unitary $S$-matrix.  $K(E)$ is a hermitian matrix in a space of two body scattering channels.  The channel momenta form a diagonal matrix, $p=\,{\rm diag}\,(\sqrt{E^{2}/4 -m_{1}^{2}}, \sqrt{E^{2}/4 -m_{2}^{2}}, ...)$.\footnote{For simplicity I have assumed equal mass particles in every channel.}  As long as $K$ is hermitian, then $S(E)$ defined by
\be
\la{kmatrix2}
S(E)=\frac{1+i\sqrt{p}K(E)\sqrt{p}}{1-i\sqrt{p}K(E)\sqrt{p}}
\ee
will be unitary.  $K(E)$ has no singularities at the channel thresholds, but on account of the form of \eq{kmatrix2}, $S(E)$ will have the correct square root branch points at every threshold.
Dalitz and Tuan parameterized $K$ as a smooth function of energy and found the pole in $\overline KN$ scattering below threshold now known as the $\Lambda(1405)$, a very interesting hadron in its own right.

Over the years a different use of the $K$-matrix has evolved.  Suppose $K$ has a pole, 
$$
K(E)=\frac{\gamma P}{E_{0}-E} + \widetilde K(E),
$$
 where $P$ is a projection matrix in the channel space ($P^{2}=P$). If $\gamma\ll 1$ and $\widetilde K(E)$ is smooth for $E\approx E_{0}$, then $S(E)$ will have a pole   below the real axis in the complex $E$-plane with imaginary part proportional to $\gamma p_{0}$, which is a resonance.  The great versatility of this parameterization is that poles in $K(E)$ can be added and $S(E)$ will always be unitary.  So here is a relatively simple way to model and fit two body scattering data:  Parameterize $K(E)$ as a sum of poles with parameters $E_{j}$,  $\gamma_{j}$, and $P_{j}$, plus a smooth (polynomial) background if desired.  Then fit the parameters to two body scattering data \cite{Anisovich:2002ij}.
 
 The obvious question is:  What is the significance of the fit parameters?  Clearly if the residue, $\gamma_{j}$, is small so the pole in the $S$-matrix is unequivocal, then the parameters of the resonance are properties of a quasi-stable eigenstate of the strong interactions and belong in the PDG tables.  This is incontrovertible.  Suppose, however, that the residue is large?  Should the existence of a pole in the $K$-matrix at energy $E_{j}$ be taken as evidence of a QCD eigenstate at that mass?  I believe the answer is definitely  ``No!''  --- a $K$-matrix pole has no \emph{a priori} claim to fundamental significance. 

The first problem is that the $K$-matrix is not unique.   There is \emph{an infinite set of equally good ``matrices'' that could have been used instead of $K$},   
\be
\la{kgeneral}
pK(\phi)=\frac{pK -\tan\phi}{1+pK \tan\phi}
\ee
where $K(0)$ is the $K$-matrix defined before\footnote{For simplicity I have ignored the matrix nature of $p$.}.  The $S$-matrix   is unitary when $K(\phi)$ is hermitian and $K(\phi)$ has the same analyticity properties as $K(0)$.  So unless one has a \emph{physical} to prefer $K$ matrix poles, any of these matrices will do.  A pole in $K(0)$ at $E_{0}$ with a small residue $\gamma_{0}$ generates as pole in $K(\phi)$ at $E(\phi)\approx E_{0}+\gamma p_{0}\tan\phi$.  So for narrow resonances the difference between $K(0)$ and $K(\phi)$ is inessential.   However, if the residue is large, the location of a pole in $K(\phi)$ will depend strongly on $\phi$.

So what was the dynamics that motivated the choice of the $K$-matrix in the first place?, or put another way, what is the significance of the parameter $\phi$ in \eq{kgeneral}?  The answer, unfortunately, is not very satisfactory and suggests that  singularities in the (usual, {\it i.\,e. $\phi=0$}) $K$-matrix have no claim to special  physical significance when their residues are large.  The physical interpretation of   $K$-matrix poles can best be understood by viewing $K$ as a limit of the   Wigner-Eisenbud $R$-matrix  developed to understand resonances in compound nuclei\cite{wignereisenbud,viki} and already mentioned in Sect.~\ref{pmatrix}.  The discussion by Blatt and Weisskopf is particularly helpful\cite{viki}, and the point can be made most simply by considering a non-relativistic, $s$-wave scattering in a single channel.  Suppose that the interaction can be ignored for $r>b$.  Then the solution to the \sch equation for $r>b$ can be parameterized   in terms of standing waves,
\be
\la{standing}
 \Psi(p,r) \propto  \frac{1}{p} \sin[p(r-b)] + R(E,b)\cos[p(r-b)]\quad\mbox{for}\ \ r\ge b
\ee
Clearly $R(E,b)$ is the inverse logarithmic derivative of $\Psi$ at $r=b$,
$$
R(E,b)=\frac{\Psi(p,b)}{\Psi'(p,b)}\  .
$$
The scattering amplitude is easily expressed in terms of $R(E,b)$,
\bea
\la{smatrix}
S = e^{2i\delta(E)} &=&e^{2ipb}\frac{1+ipR(E,b)}{1-ipR(E,b)}\quad\mbox{or} \nonumber \\
\tan(\delta(E)-pb)&=&pR(E,b)
\eea
It is easy to show that at low energies, when $p\ll 1/b$, there is negligible scattering \emph{except at energies, $E_{j}$, where $\Psi'(E_{j},b)\approx 0$}.  As the energy rises through $E_{j}$, the phase shift rises through $\pi/2$ and the scattering resonates.  So, the problem of finding resonances can be reduced to solving a \emph{stationary state} \sch equation, $H\Psi=E\Psi$ for $r\le b$, subject to the unusual boundary condition, $\Psi'=0$ at $r=b$.

In the limit that the size of the scattering region (in Wigner's case, the nucleus) is negligible compared to the de Broglie wavelength, {\it i.\,e.} when $p\ll 1/b$, then $b$ can be neglected throughout:  So $R(E,b)\to R(E,0)$, and 
\be 
\la{rtokmatrix}
S =  \frac{1+ipR(E,0)}{1-ipR(E,0)}
\ee
Comparing with \eq{kmatrix2}, it is clear that \emph{the $K$-matrix is the $b\to 0$ limit of Wigner's $R$-matrix}: 
$$
K(E)=R(E,0)
\quad \mbox{and}\quad K(E)=\frac{1}{p}\tan\delta(E)
$$
At a pole in the $K$-matrix the scattering wavefunction emerges from the origin already phase shifted by $(n+\half)\pi$, corresponding to a resonant system whose dimensions are negligibly small compared to the deBroglie wavelength of the scattering particles.  This is an excellent approximation for the case of interest to Wigner:  thermal neutrons scattering off nuclei.
 It may have seemed a reasonable starting point for an attempt to describe resonances back in the 1950's when it was not clear that hadrons had complex internal structure.  However it certainly does not look hadron resonances as now understood in QCD, where the typical interaction scale is of order 1 fermi and the typical center of mass momentum considerably greater than 1 fermi$^{-1}$.  

With this discussion in mind let us return to the ambiguity parameterized by $\phi$ in \eq{kgeneral}.  In terms of the phase shift,
$K(\phi)=\frac{1}{p}\tan(\delta(E)-\phi)$, so a pole in $K(\phi)$ occurs when $\delta(E)=(n+1/2)\pi+\phi$.  Near a \emph{narrow} resonance the phase shift rises rapidly through $\pi$, so no matter what $\phi$ one preferred, there would be a pole in $K(\phi)$ near the place where the physical phase shift varied dramatically.  Quibbling about $\phi$ corresponds to debating the precise definition of the resonant energy.  The ambiguity is only important for broad enhancements, and the lesson is that the energy at which the $K$-matrix has a pole has no special significance in QCD.  The larger the residue the less meaningful the location of a $K$-matrix pole. 

 The deep question lurking behind this whole discussion is:  What is the physically appropriate framework that can connect QCD calculations, either numerical or phenomenological, with the analysis of scattering data.  The challenge to find such a framework is made more pressing by the realization that extraordinary hadrons exist, are interesting, and need to be understood.
 
\section{Acknowledgments}

I would like to thank Jose Pel\'aez for correspondence and the Durham Institute for Particle Physics Phenomenology for its hospitality during the summer of 2006 when this work was begun.
This work is supported in part by funds provided by the
U.S.~Department of Energy (D.O.E.) under cooperative research
agreement DE-FC02-94ER40818.

 \bibliographystyle{aipprocl}   

\begin{thebibliography}{99}
\bibitem{dalitzplot}
R.~H.~Dalitz, Phil.\ Mag.\ {\bf 44}, 1068 (1953).
\bibitem{Castillejo:1955ed}
  L.~Castillejo, R.~H.~Dalitz and F.~J.~Dyson,
  Phys.\ Rev.\  {\bf 101}, 453 (1956).
\bibitem{Dalitz:1960du}
  R.~H.~Dalitz and S.~F.~Tuan,
  Phys.\ Rev.\ Lett.\  {\bf 2}, 425 (1959).
  R.~H.~Dalitz and S.~F.~Tuan,
  Annals Phys.\  {\bf 8}, 100 (1959);  {\bf 10}, 307 (1960).
\bibitem{Caprini:2005an}
  I.~Caprini, G.~Colangelo and H.~Leutwyler,
  Int.\ J.\ Mod.\ Phys.\ A {\bf 21}, 954 (2006)
  [arXiv:hep-ph/0509266];
  Phys.\ Rev.\ Lett.\  {\bf 96}, 132001 (2006)
  [arXiv:hep-ph/0512364];
  H.~Leutwyler,
  arXiv:hep-ph/0608218.
\bibitem{kappa} For a discussion with references, see the S.~Spanier and N.~Tornqvist,  ``Note on the Scalar Mesons'' in Ref.~\cite{Yao:2006px}.
\bibitem{Yao:2006px}
  W.~M.~Yao {\it et al.}  [Particle Data Group],
  J.\ Phys.\ G {\bf 33}, 1 (2006). 
\bibitem{Rosner:2006jz} For a recent review with extensive references, see 
  J.~L.~Rosner,
  arXiv:hep-ph/0609195.
  \bibitem{Close:2002zu}
  F.~E.~Close and N.~A.~Tornqvist,
  J.\ Phys.\ G {\bf 28}, R249 (2002)
  [arXiv:hep-ph/0204205].
\bibitem{Amsler:2004ps}
  C.~Amsler and N.~A.~Tornqvist,
  Phys.\ Rept.\  {\bf 389}, 61 (2004).
  \bibitem{Achasov:2006sr}
  N.~N.~Achasov,
  arXiv:hep-ph/0609261.
%
\bibitem{vanBeveren:2006ua}
  E.~van Beveren, D.~V.~Bugg, F.~Kleefeld and G.~Rupp,
  Phys.\ Lett.\ B {\bf 641}, 265 (2006)
  [arXiv:hep-ph/0606022].
  \bibitem{Anisovich:2005kt}
  A.~V.~Anisovich, V.~V.~Anisovich, L.~G.~Dakhno, V.~N.~Markov, M.~A.~Matveev, V.~A.~Nikonov and A.~V.~Sarantsev,
  arXiv:hep-ph/0508260.
\bibitem{Pelaez:2003dy} 
  J.~R.~Pelaez,
  AIP Conf.\ Proc.\  {\bf 688}, 45 (2004)
  [arXiv:hep-ph/0307018];
  J.~R.~Pel\'aez,
  Phys.\ Rev.\ Lett.\  {\bf 92}, 102001 (2004)
  [arXiv:hep-ph/0309292];
  AIP Conf.\ Proc.\  {\bf 814}, 670 (2006)
  [arXiv:hep-ph/0510118].

 \bibitem{chewmandelstam}  G. F. Chew and S.  Mandelstam, Phys.\ Rev.\ {\bf 119}, 467 (1960).
 \bibitem{frautschi}  For an introduction to the $N/D$ method see S.\ Frautschi, \emph{Regge Poles and S-Matrix Theory}, (W. A. Benjamin, New York, 1963).
 \bibitem{diquarks} For a review an further references, M.~Anselmino, E.~Predazzi, S.~Ekelin, S.~Fredriksson and D.~B.~Lichtenberg, Rev.
Mod. Phys. {\bf 65}, 1199 (1993), or M.~Anselmino, E.~Predazzi, eds. {\sl International Workshop on Diquarks and Other Models
of Compositeness: Diquarks III, Turin, Italy, 28-30 Oct 1996} (World Scientific, 1998).  For a recent review see, 
  R.~L.~Jaffe,
  Phys.\ Rept.\  {\bf 409}, 1 (2005)
  [Nucl.\ Phys.\ Proc.\ Suppl.\  {\bf 142}, 343 (2005)]
  [arXiv:hep-ph/0409065].
  \bibitem{chiral} A. A. Andrianov, Phys.\ Lett.\ {\bf B 157}, 425 (1985). A. A. Andrianov and L. Bonora, Nucl.\ Phys.\ {\bf B 233},
232 (1984). D. Espriu, E. de Rafael and J. Taron, Nucl. Phys. {\bf B 345}, 22 (1990); S. Peris and E. de
Rafael, Phys.\ Lett.\ {\bf B 348}, 539 (1995).
\bibitem{feshfano} They are named after Herman   Feshbach, who independently discovered and applied the idea  with great
success in nuclear physics\cite{feshbach}.  The basic ideas were developed in the early days of quantum mechanics by Rice\cite{rice}, and especially by Fano\cite{fano}, who applied them to problems in  atomic physics.
\bibitem{feshbach}H.~Feshbach, Ann. Phys. (N.Y.) 5, 357 (1958).
\bibitem{rice} O.~K.~Rice, J.~Chem.~Phys.~{\bf 1}, 375 (1933).
\bibitem{fano} U.~Fano, Nuovo Cimento {\bf 12}, 154 (1935); Phys. Rev. {\bf 124},
1866 (1961).
\bibitem{Jaffe:1981up}
  R.~L.~Jaffe,
MIT-CTP-951
{\it Rapporteur's talk presented at Lepton Photon Symp., Bonn, Germany, Aug 24-29, 1981}, (Scanned version available at KEK.)

 \bibitem{'tHooft:1973jz}
  G.~'t Hooft,
  Nucl.\ Phys.\ B {\bf 72}, 461 (1974).
  \bibitem{Coleman:1980nk}
  S.~R.~Coleman, {\sl $1/N$}
  in  {\sl Aspects of Symmetry}
(Cambridge, 1985). 
\bibitem{Gasser:1983yg}
  J.~Gasser and H.~Leutwyler,
  Annals Phys.\  {\bf 158}, 142 (1984).
J. Gasser and H. Leutwyler, Annals Phys.\ {\bf 158}, 142 (1984). Nucl.\ Phys.\ {\bf B 250} (1985) 465.  

\bibitem{truong} T.~N.~Truong, Phys.\ Rev.\ Lett.\ {\bf 61}, 2526 (1988).
\bibitem{iam}A.~Dobado, M.~J.~Herrero, and T.~N.~Truong, Phys.\ Lett.\ {\bf B 235}, 134 (1990); T.~N.~Truong, Phys.\ Rev.\ Lett.\ {\bf 67}, 2260 (1991); A.~Dobado and J.~R.~Pel«aez, Phys.\  Rev.\ {\bf D
47}, 4883 (1993), Z.\ Phys.\  {\bf C 57}, 501 (1993), [arXiv:hep-ph/9604416];
 J.A. Oller, E. Oset, and J.R. Pel\'aez, Phys. Rev. {\bf D59}~(1999),  
 074001, 
 Erratum Phys. Rev. {\bf D60}(1999), 09906; 
 Phys. Rev. {\bf D62}~(2000),114017;
  A.~Gomez Nicola and J.~R.~Pel\'aez,
  Phys.\ Rev.\ D {\bf 65}, 054009 (2002)
  [arXiv:hep-ph/0109056].
  

\bibitem{Pelaez:2006nj}
  J.~R.~Pelaez and G.~Rios,
  arXiv:hep-ph/0610397.
\bibitem{Uehara:2003ax}
M. Uehara, [arXiv:hep-ph/0308241]. [arXiv:hep-ph/0401037]. [arXiv:hep-ph/0404221].

\bibitem{Pelaez:2006qv}
  J.~R.~Pelaez,
  arXiv:hep-ph/0612052.
  
\bibitem{Rosner:2006vc}
  J.~L.~Rosner,
  arXiv:hep-ph/0608102.
\bibitem{Jaffe:1990xj}
  R.~L.~Jaffe,
  Phys.\ Lett.\ B {\bf 245}, 221 (1990);
  R.~L.~Jaffe and P.~F.~Mende,
  Nucl.\ Phys.\ B {\bf 369}, 189 (1992).
\bibitem{Jaffe:2004ph}
  R.~L.~Jaffe,
  Phys.\ Rept.\  {\bf 409}, 1 (2005)
  [Nucl.\ Phys.\ Proc.\ Suppl.\  {\bf 142}, 343 (2005)]
  [arXiv:hep-ph/0409065].
  \bibitem{Jaffe:1976ig}
  R.~L.~Jaffe,
  Phys.\ Rev.\ D {\bf 15}, 267, 281 (1977).
\bibitem{Jaffe:1978bu}
  R.~L.~Jaffe and F.~E.~Low,
  Phys.\ Rev.\ D {\bf 19}, 2105 (1979),
  F.~E.~Low,
   ``Quark Model States And Low-Energy Scattering,''
  {\it Pointlike Structures Inside and Outside Hadrons: Proceedings of the 17th International School of Subnuclear Physics,  Erice, Italy}, A.~Zichichi, ed. (Plenum, NY, 1979),
  R.~L.~Jaffe,
  ``How To Analyze Low-Energy Scattering,'' in
{\it Asymptotic Realms Of Physics} (MIT Press, Cambridge, 1982).


\bibitem{wignereisenbud} E.~P.~Wigner and L.~Eisenbud, Phys.\ Rev.\ {\bf 72} 29 (1947); G.~Breit and W.~G.~Bouricius, Phys.\ Rev.\ {\bf 75}, 1029 (1949); H.~Feshbach and E.~L.~Lomon, Ann.\ Phys. (N.~Y.) {\bf 29}, 19 (1964).
\bibitem{Lenz:1985jk}
  F.~Lenz, J.~T.~Londergan, E.~J.~Moniz, R.~Rosenfelder, M.~Stingl and K.~Yazaki,
  Annals Phys.\  {\bf 170}, 65 (1986).
\bibitem{Dalitz:1959dn}
  R.~H.~Dalitz and S.~F.~Tuan,
  Phys.\ Rev.\ Lett.\  {\bf 2}, 425 (1959),
  Annals Phys.\  {\bf 10}, 307 (1960).
\bibitem{Anisovich:2002ij} See, for example,
  V.~V.~Anisovich and A.~V.~Sarantsev,
  Eur.\ Phys.\ J.\ A {\bf 16}, 229 (2003)
  [arXiv:hep-ph/0204328], and further references therein.
\bibitem{viki}J.~Blatt and V.~F.~Weisskopf, {\sl Theoretical Nuclear Physics} (Springer-Verlag, New York, 1952), pp 544--549.
\end{thebibliography}



\end{document}